\documentclass[aps,twocolumn,groupedaddress,superscriptaddress]{revtex4}%

\usepackage{amsfonts}
\usepackage{amsmath}
\usepackage{amssymb}
\usepackage{graphicx}
\usepackage{float}

\providecommand{\U}[1]{\protect\rule{.1in}{.1in}}
\begin{document}

\title{Spin-charge interplay in antiferromagnetic LSCO studied by the muons,
neutrons, and ARPES techniques}
\author{Gil Drachuck}
\affiliation{Department of Physics, Technion - Israel Institute of Technology, Haifa,
32000, Israel}
\author{Elia Razzoli}
\affiliation{Swiss Light Source, Paul Scherrer Institute, CH-5232 Villigen PSI,
Switzerland}
\affiliation{Département de Physique and Fribourg Center for Nanomaterials, Université de Fribourg, CH-1700 Fribourg, Switzerland}
\author{Galina Bazalitski}
\affiliation{Department of Physics, Technion - Israel Institute of Technology, Haifa,
32000, Israel}
\author{Amit Kanigel}
\affiliation{Department of Physics, Technion - Israel Institute of Technology, Haifa,
32000, Israel}
\author{Christof Niedermayer}
\affiliation{Laboratory for Neutron Scattering, Paul Scherrer Institute, CH-5232 Villigen
PSI, Switzerland}
\author{Ming Shi}
\affiliation{Swiss Light Source, Paul Scherrer Institute, CH-5232 Villigen PSI,
Switzerland}
\author{Amit Keren}
\affiliation{Department of Physics, Technion - Israel Institute of Technology, Haifa,
32000, Israel}
\date{\today }
\maketitle



Exploring whether a spin density wave (SDW) is responsible for the charge
excitations gap in the high-temperature superconducting cuprates is
difficult, since the region of the phase diagram where the magnetic
properties are clearly exposed is different from the region where the band
dispersion is visible. On the one hand, long range magnetic order disappears
as doping approaches 2\% from below, hindering our ability to perform
elastic neutron scattering (ENS). On the other hand, cuprates become
insulating at low temperature when the doping approaches 2\% from above,
thus restricting angle-resolved photoemission spectroscopy (ARPES). In fact,
ARPES data for samples with doping lower than 3\% are rare and missing the
quasiparticle peaks in the energy distribution curves (EDCs) \cite{ShenPRB04}
\cite{VallaJPCM12}. The main problem is the high resistivity of extremely
underdoped samples, which is detrimental to ARPES due to charging effects.
Nevertheless, the resistivity of La$_{2-x}$Sr$_{x}$CuO$_{4}$ [LSCO] as a
function of temperature, at 2\% doping, has a broad minimum around 100~K
\cite{AndoPRL04}. This minimum opens a window for both experiments. By
preparing a series of LSCO single crystals with $\sim $0.2-0.3\% doping
steps around 2\%, we managed to find one to which both techniques apply.
This allows us to explore the cross talk between the magnetic and electronic
properties of the material.

\begin{figure*}[hbtp]
\begin{center}
\includegraphics[trim=0cm 5cm 0cm 5cm ,clip=true,width=15cm]{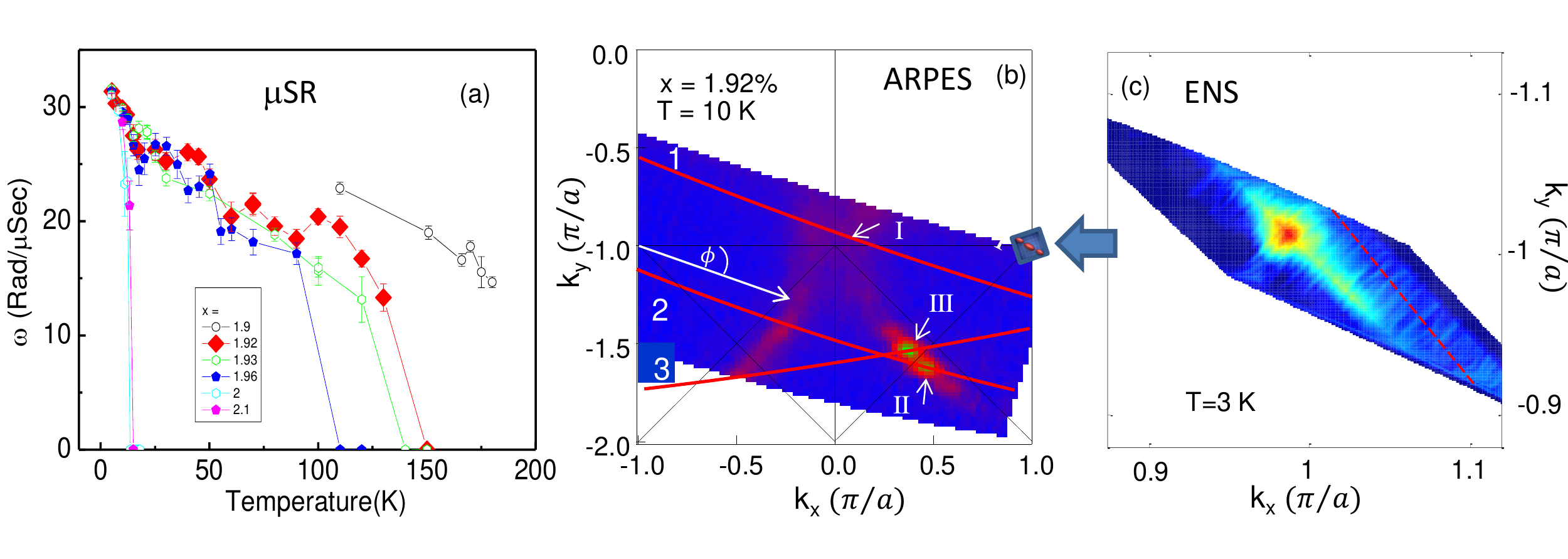}
\end{center}
\caption{(a) muon spin rotation frequency as a function of temperature for
LSCO single crystals with different $x$ values. At $x=2\%$ there is a clear
transition from an antiferromagnetic to a spin-galss ground states. The data
in the rest of this paper is from the x=1.92\% sample which is in the
antiferromangetic phase. (b) Low energy ARPES data in the second Brillion
Zone. The red lines 1, 2 show cuts on which data was collected. Line 3 is
the cut on which temperauter dependence data is presented in Fig. 2. Point
I, II, and II mark points where EDCs are extracted. The intensity is
centered in the diagonal region. The inset at $\mathbf{k}=(1,-1)$ is an
illustration of the location of the commensurate and incommensurate magnetic
peaks. (c) Elastic neutron scattering measurements at 3~K around a magnetic
reciprocal lattice vector showing two incommensurate peaks. The red lines
depicts the measurements scan direction.}
\label{Fig1}
\end{figure*}

The series of samples are first characterized by muon spin rotation ($\mu $%
SR). In Fig.~1a we show the muon spin rotation frequency as a function of
temperature for all the crystals. In samples with doping equal to or less
than 2\%, the oscillation starts at temperatures on the order of 100~K. At
2\%, there is a sharp transition from the antiferromagnetic long range order
to the spin glass state where the same oscillations appear only at
temperatures on the order of 10~K \cite{NeidermayerPRL98}. More details of
the $\mu $SR characterization of the samples are given below and in the
supplementary material. We perform the ENS and ARPES measurements on a
sample with $x=1.92$\% (red rhombuses) which is antiferromagnetic with a N%
\'{e}el temperature $T_{N}=140$~K. Fig. 1b shows the ARPES intensity map
near the Fermi level ($E_{F}$) in the second Brillouin Zone (BZ), measured
along cuts parallel to cut 1 and 2 shown in the figure. The high intensities
represent the underlying Fermi surface (FS) and has the morphology of the FS
in more doped LSCO \cite{RazzoliNJP} and other cuprates. Most of the
intensity is centered around the zone diagonal. This suggests the presence
of gapped electronic excitations in the off-diagonal $(0,\pi )$ region.
Similar results were obtained at a higher doping level of $x=3$ -- $8\%$
\cite{Razzoli}.

Figure 1c depicts ENS for the same sample at 3~K. The scans are performed in
a narrow range in the reciprocal lattice space centered around the $(1,-1)$
point, in the standard tetragonal ARPES units of $\pi /a$. The scanned area
of the ENS experiment is presented on the BZ of the ARPES experiment by the
small inset in Fig. 1b. Two incommensurate peaks are present in the center
of the figure. Such incommensurate peaks occur when, on top of the main
magnetic order, the system develops spin modulations (stripes) running in
diagonal to the bond directions (diagonal stripes). The stripes and
commensurate scattering observed in our sample are in agreement with
previous reports for low doping LSCO \cite{FujitaPRB02,MatsudaPRB2002}, and
are discussed further below.

\begin{figure*}[]
\begin{center}
\includegraphics[trim=0cm 0cm 0cm 0cm, width=15cm]{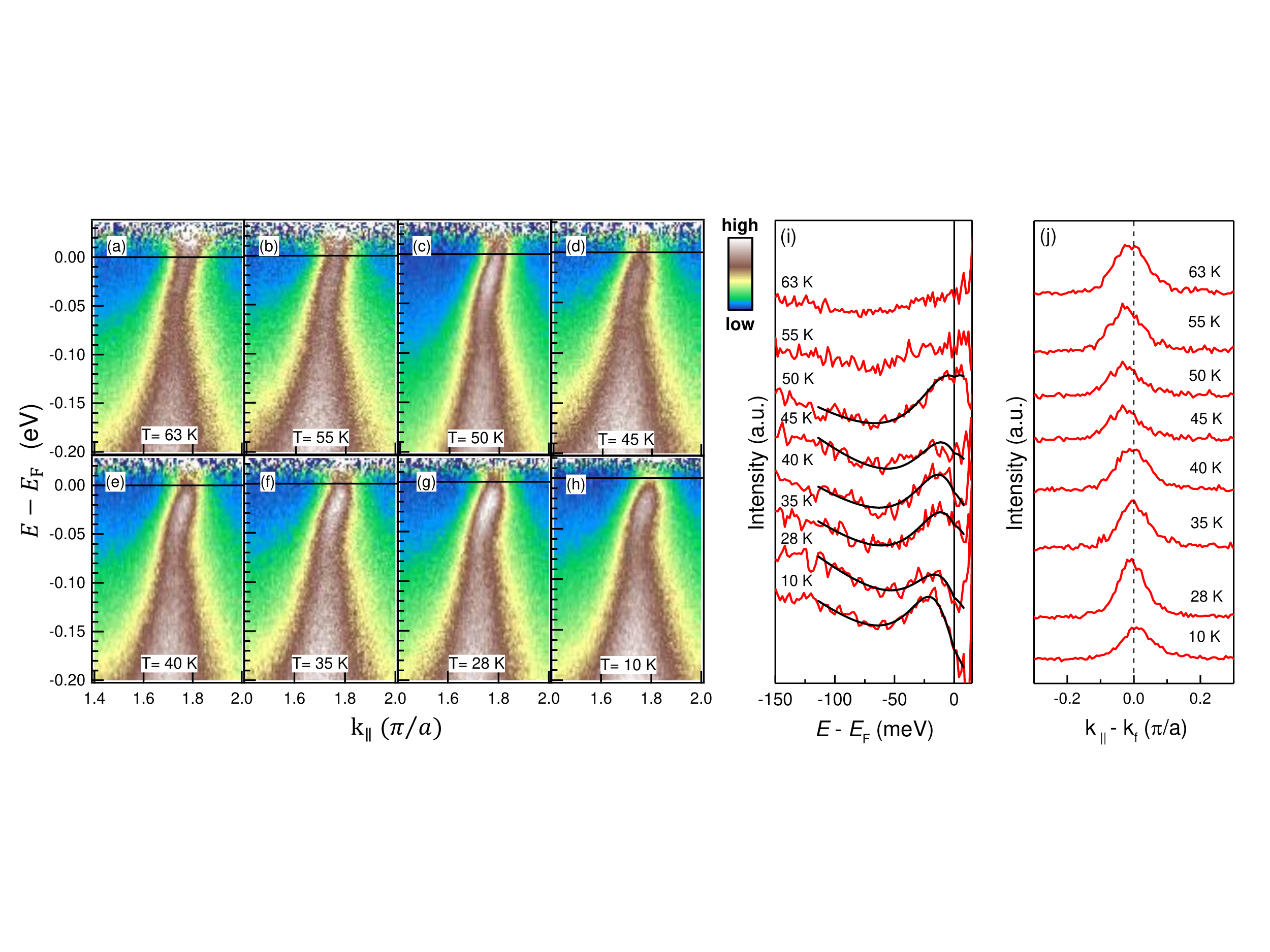}
\end{center}
\caption{(a-h) ARPES spectra along a cut given by line 3 in Fig. 1. The
spectra are divided resoluiton broadened Fermi-Dirac distribution. At high $%
T $ the spectrum crosses the Fermi energy. At low $T$ it does not. (i) EDCs
at the Fermi momentum of point III in Fig. 1b for different temperatures.
The solid lines are guides to the eye. At 10K the peak is clearly below the
Fermi energy. $\Delta(T)$ is presented in Fig. 4c. (j) MDC at $E_{F}$ as a
function of temperature showing that $k_{F}$ in the diagonal region is
temperature independent.}
\label{Fig2}
\end{figure*}

In Fig. 2a-h, we show the temperature dependence of ARPES data in the
vicinity of point III along cut 3 in Fig. 1b. All spectra are divided by
resolution broadened Fermi-Dirac distribution at the nominal temperature.
Data analysis by the Lucy--Richardson deconvoltion method \cite%
{LRDeconvolution} is discussed in the supplementary material. At high
temperatures ($T>50$~K), the dispersive peak crosses $E_{F}$ as clearly seen
in Fig. 2a. In contrast, at low temperatures ($T<45$ K) there is a gap in
the electronic spectra, \emph{i.e.} the peak position at $k_{F}$ stays below
$E_{F}$. A similar gap exists in LSCO up to 8\% doping \cite{Razzoli}, in
LBCO at 4\% \cite{VallaJPCM12}, in the electron doped compound \cite%
{HarterPRL12}, and in Bi2212 \cite{VishikPNAS12}, and also in
simulations where inhomogeneous SDW and superconductivity coexist \cite%
{AtkinsonXXX}. EDCs extracted from Fig. 2a-h, at $k_{F}$ given by point III
of Fig. 1b, are plotted in Fig. 2i. The solid lines are guides to the eye.
At temperatures above 45~K, the peak is at $E_{F}$. However, below this
temperature the peaks are clearly below $E_{F}$, indicating the presence of
a gap. Another important aspect of these EDCs is the fact that at all low
temperatures, $E_{F}$ is not in the middle of the gap, namely, the intensity
of the spectra immediately above $E_{F}$ is lower than at $E_{F}$. This
could be an analysis artifact due to the division by the resolution
broadened Fermi-Dirac function \cite{RameauJESRP10} or indicate that the
particle-hole symmetry is broken. Particle-hole asymmetry toward the
diagonal region in the non-superconducting phase \cite{YangNature08}, and in
the off-diagonal region \cite{HashimotoNatPhys10}, was found previously.
However, it is very difficult to distinguish between the two options in our
sample with extremely low carrier concentration.

Another parameter relevant for understanding the mechanism by which this
diagonal gap opens is the temperature dependence of $k_{F}$. This is
explored by extracting the momentum dispersion curve (MDC) at $E_{F}$ from
Fig. 2a-h, as presented in Fig.~2j. The peak position fluctuates somewhat
around a constant value as the temperature is lowered, indicating that $%
k_{F}=0.63(\pi /a)$ along the diagonal is temperature-independent.
Therefore, the high intensity curves in Fig.~1b represents the FS.

The evolution of the gap around the FS is shown in Fig. 3a-b for 10 and
100~K. The spectra marked I and II in Fig.~3a are EDCs at the $\mathbf{k}%
_{F} $ of points I and II in Fig. 1b. The spectrum in between correspond to $%
\mathbf{k}_{F}$ between point I and II. EDCs shown in Fig. 3b are from
similar point along the FS. In the off-diagonal region, there is no spectral
weight at $E_{F}$ for both temperatures, indicating that the off-diagonal
gap opens at a temperature higher than 100~K. In the diagonal region, there
is high spectral weight at 100~K, but not at 10~K. Thus at 100~K we observe
a Fermi arc \cite{KanigelNatPhys06}, while at 10~K a gap appears all around
the FS.

\begin{figure}[tbph]
\begin{center}
\includegraphics[trim=5cm 0cm 5cm 0cm, height=10cm]{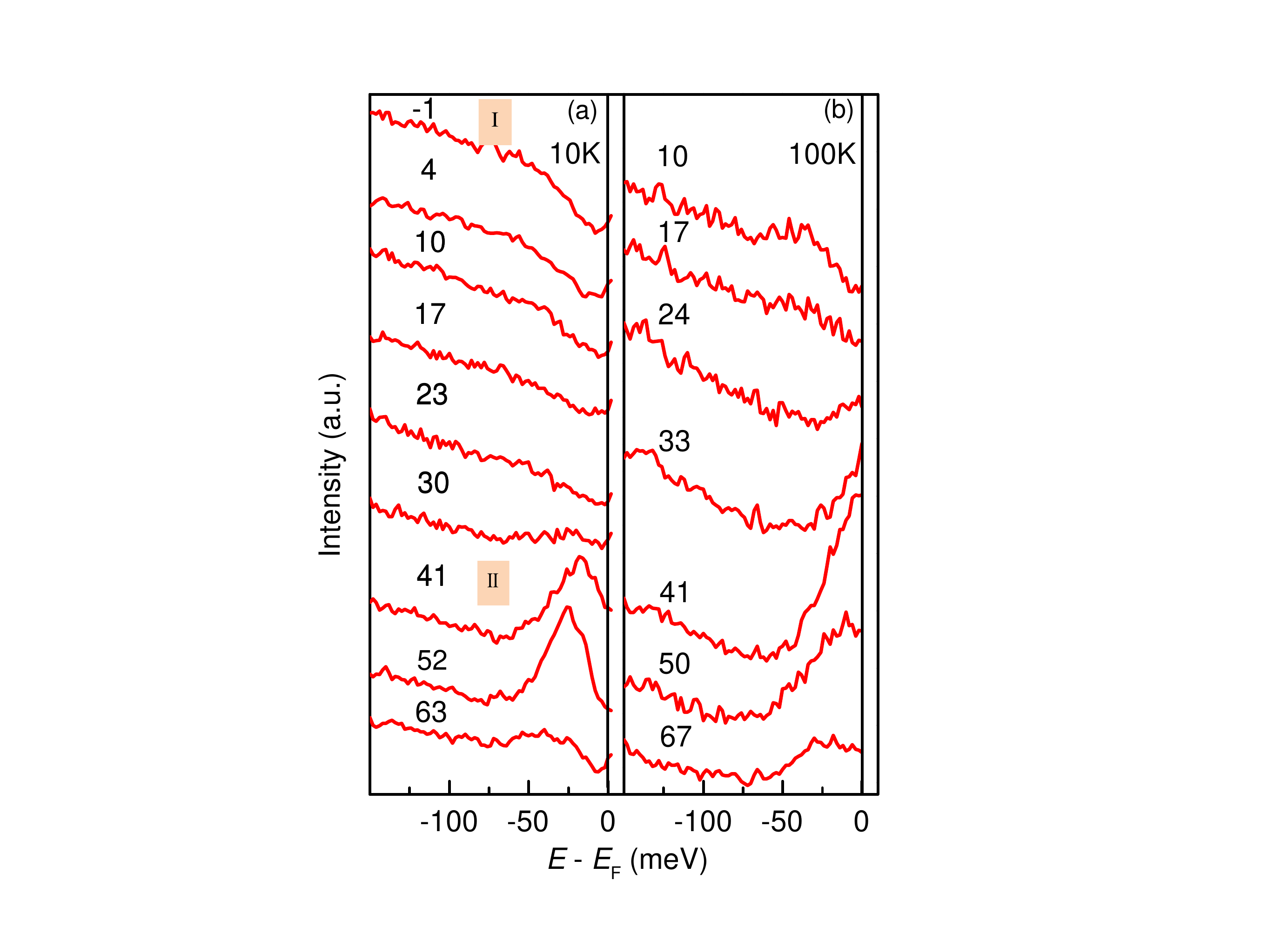}
\end{center}
\caption{(a) EDCs at different $\mathbf{k}_{F}$ along the FS in two
different temperatures. The fermi surface angle $\protect\phi$ is indicated
above each EDC. Spectra I and II are taken at the $\mathbf{k}$ of point I
and II in Fig. 1b which are defined by the crossing point between the red
lines 1 and 2 and the FS. The other EDCs are from the crossing points in
between. At 10K a gap is observed all around the FS. $\Delta(\protect\phi)$
is presented in the inset of Fig. 4c. (b) Similar EDCs at 100~K. In this
case a gap is observed only in the off diagonal region.}
\label{Fig3}
\end{figure}

Finally, in Fig. 4 we present a summary of parameters extracted from all
three techniques. Fig.~4a shows the temperature dependence of the muon
rotation frequency and magnetic volume fraction. The inset represents a fit
of a three-component model, with one frequency, to the $\mu $SR data at
different temperatures. More details are available in the supplementary
material. The magnetic parameters grow upon cooling, saturate between $50$
and $25$~K, and grow again at lower temperatures. The temperature $T_{f}=25$%
~K was associated with hole freezing into one dimensional domain walls \cite%
{NeidermayerPRL98,ChargeFreezing}. The restriction of the charge motion
below $T_{f}$ must be intimately related to the opening of the nodal gap.

Fig.~4b shows the commensurate and the incommensurate scattering intensity.
The inset shows the raw data along the red diagonal line in Fig. 1c (in
orthorhombic units). As the temperature is raised, the incommensurate
intensity decreases until it disappears at $T_{f}=30$~K. The commensurate
intensity peaks at $25~K$, just when the muon oscillation change its
behavior, and vanishes at $T_{N}=130$~K. Our ENS temperature dependence
results and incommensurability parameter (in tetragonal units) $\delta
=0.027(\pi /a)$ are in good agreement with previously reported data~\cite%
{MatsudaPRB2002}. While the ENS data is well understood in terms of a spiral
ground state \cite{LuscherPRL07}, its relation to the gap is still
mysterious, since the nesting condition $k_{F}+\delta =\frac{\sqrt{2}\pi }{2a}$ is not fulfilled in our sample.

The diagonal gaps at $k_{F}$, $\Delta $, are determined from the peak in the
EDCs of Fig.~2i and presented in Fig.~4c as a function of temperature. Very
similar gap values are found using the Lucy--Richardson de-convolutions
method (see supplementary material) \cite{LRDeconvolution}. The gap opens at
$50$~K. The major observations in this figure are: I) the diagonal gap opens
only when the commensurate moment is nearly at its full value. II) the
diagonal gap opens when the incommensurate moment is not detectable by ENS.
The angular dependence of the gaps is presented in the inset of Fig.~4c; the
angle $\phi $ is defined in Fig. 1b,. When a peak is absent, the gap is
taken from the change of slope in the EDCs. $\Delta (\phi )$ is in agreement
with previous measurements for 3 to 8\% dopings \cite{Razzoli}.

\begin{figure}[h]
\begin{center}
\includegraphics[trim=7cm 0cm 5cm 0cm, clip=true,height=11cm]{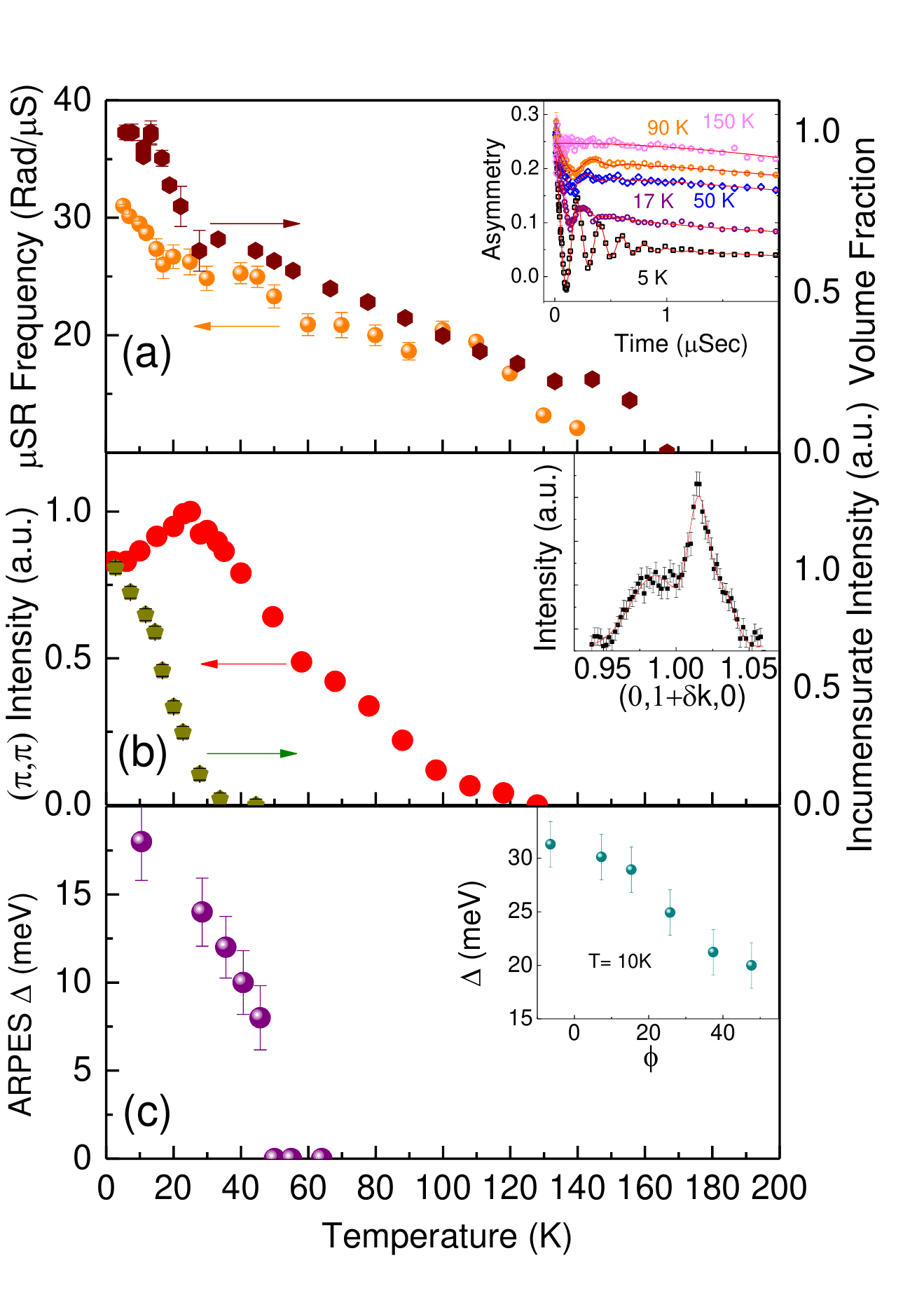}
\end{center}
\caption{A summary of experimental parameters from all three techniques. (a)
Muon rotation frequency and the magnetic volume fraction of the sample as a
function of temperature taken from a fit to a three Lorenzians model (see
supplementary material). The inset shows raw data and the fit quality for
the x=1.92\% sample. (b) Commensurate and incommensurate elastic neutron
scattering intensity. The inset is the intensity versus $k$ along the
neutron scattering cut in Fig. 1c. (c) ARPES diagonal gap $\Delta $ at $%
k_{F} $, as obtained from the peaks in Fig. 2i, versus temperature. The
inset shows the angular dependence of the gap $\Delta (\protect\phi )$.}
\label{Fig4}
\end{figure}

On the basis of weak coupling theory, Berg \emph{et al.} \cite{BergPRL08}
argued that the commensurate part of the SDW can open a diagonal gap only if
the moment is larger than a critical value. However, the opening of the gap
will be accompanied by a shift in $k_{F}$. The effect of the incommensurate
part of the SDW depends on the properties of time reversal followed by
translation symmetry, which can be broken or unbroken. In the unbroken case
a nodal gap will open when the perturbation exceeds a critical value. Both
these options do not agree with our data. \ In the broken symmetry case, a
nodal gap can open for arbitrarily small perturbation with no impact on $%
k_{F}$. This option is in agrement with our measurements and bares important
information on the symmetries of the ground state. Another possibility is
that the gap is of the superconducting type, which maintains particle-hole
symmetry and keeps $k_{F}$ fixed. This option would mean that the cuprates
have a superconducting gap in the AFM phase.

To summarize, we detect a nodal gap in the AFM phase of La$_{2-x}$Sr$_{x}$CuO%
$_{4}$. The gap opens well below $T_{N}$ and a bit above the temperature
where incommensurate SDW is detected. This finding puts strong restrictions
on the origin of the nodal gap.
\section{Supplementary Material}

\subsection{$\protect\mu$SR}

The experiment was carried out at PSI on the GPS beam line. We fit the muon
polarization as a function of time to
\begin{equation*}
\begin{array}{l}
{P_z}\left( t \right) = \left( {1 - {V_m}} \right){{\mathop{\rm e}\nolimits} ^{ - \frac{1}{2}{D^2}{%
t^2}}} + \\
{V_m}\left[ {p \times {{\mathop{\rm e}\nolimits} ^{ - {{\left( {{R_1}t}
\right)}^{{b_1}}}}} + \left( {1 - p} \right) \times {{\mathop{\rm e}\nolimits%
} ^{ - {{\left( {{R_2}t} \right)}^{{b_2}}}}}\cos \left( {\omega t + \phi }
\right)} \right] + {P_{Bg}}%
\end{array}%
\end{equation*}%
where $\omega $ is the muon rotation angular frequency, $V_{m}$ stands for
the magnetic volume fraction, $p$ is the amplitude of the rotating signal
resulting from the angle between the muon spin and the internal field, $D,$ $%
R_{1}$ and $R_{2}$ are relaxation rates, $b_{1}$ and $b_{2}$ are stretching
exponents, and $P_{Bg}$ is the background polarization from muon that missed
the sample. We could fit the data well with $b_{1}=0.5$ and $b_{2}=1$ for
most of the temperatures. Between $T=25$~K to 50~K we allowed freedom in
their values to achieve an optimal fit.

\subsection{ARPES}

The experiment was done in PSI using conditions similar to those in Ref.
\cite{ShiPRL08}. Charging tests have been carried out by varying the photon
flux. No charging was found. In addition, after performing the high
temperature measurements, the sample was cooled again and the results at
10~K were reproducible. We examined the surface with LEED at each
temperature and no degeneration or reconstruction was found during the
measurements.

\begin{figure}[]
\begin{center}
\includegraphics[height=8cm]{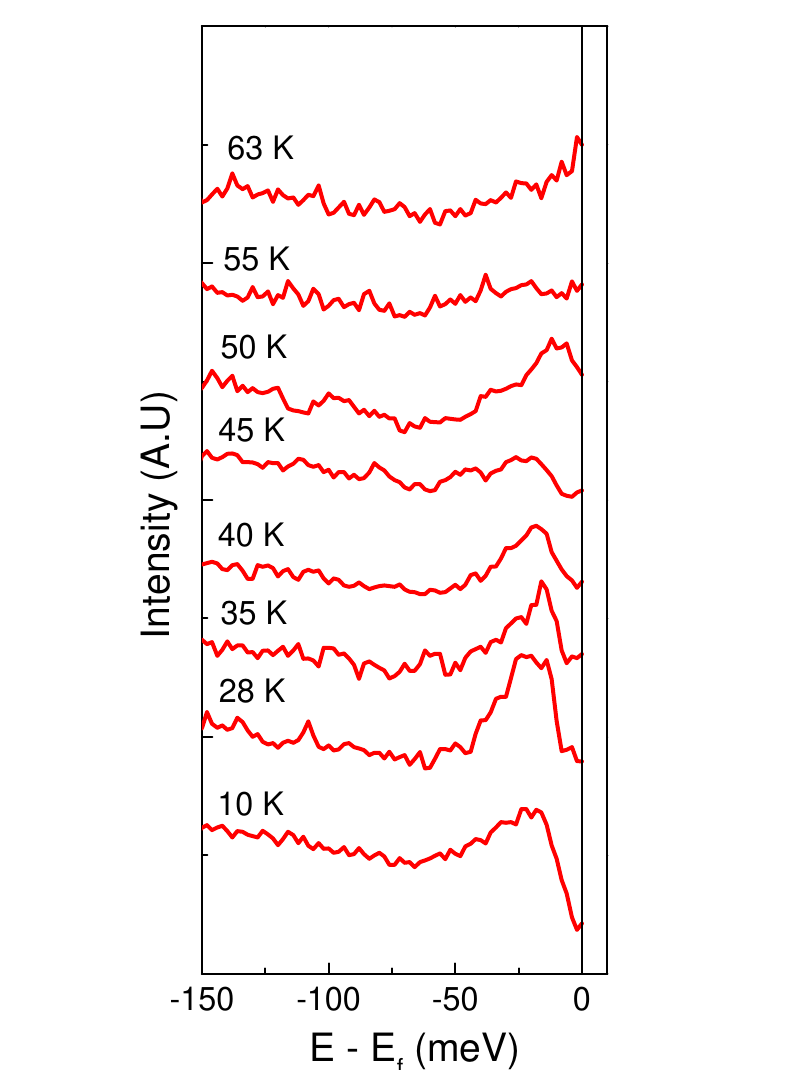}
\end{center}
\caption{The same data as in Fig. 2i but analyzed using the Lucy-Richardson
deconvoltion method \protect\cite{LRDeconvolution}.}
\label{Fig5}
\end{figure}

It should be pointed out that in ARPES the incoming light warms the surface
and there could be a few degrees difference between the temperature of the
surface from which electrons are ejected and the temperature of the cold
figure which we quote in the paper.

EDCs obtained by LRM are depicted in Fig.~\ref{Fig5}.

\subsection{Neutron diffraction}

The neutron diffraction experiments was carried out at Paul Scherrer
Institute (SINQ, PSI). The crystal was mounted on an aluminium sample holder
with the $\mathbf{\hat{c}}$ perpendicular to the neutron beam; $%
(1,0,0)-(0,1,0)$ scattering plane. The collimation was 80'-40'-80'.
Beryllium filter was placed before the analyzer to remove higher order
neutrons. The instrument has a nine-blade pyrolytic graphite (PG) analyzers.
The diffraction studies were performed with 4.04A (5 meV) neutrons, using
all blades to create a 2D color plot. The sample was aligned in such a
manner that the area of interest (commensurate and incommensurate magnetism)
was measured in the center blade. The incommensurate part was measured in a $%
(0,1+\delta q,0)$ scan direction. The high temperature data was subtracted
to remove the $\lambda /2$ contribution form the $(2,0,0)$ structural Bragg
peak. For the commensurate part, the measurement was performed with a scan
direction $(\delta q,1+\delta q,0)$ to avoid the incommensurate contribution.

\section{Acknowledgements}

The Technion team was supported by the Israeli Science Foundation (ISF) and the
joint German-Israeli DIP project. E.R. was partially supported by the Fonds
National Suisse pour la Recherche Scientifique through Div. II and the Swiss
National Center of Competence in Research MaNEP. We thank the SLS, S$\mu$S and SINQ
beam line staff at the Paul Scherrer Institute for their excellent support.

\end{document}